\documentclass[%
 sor,
twoside,
 jor,
 amsmath,amssymb,
 reprint,%
author-numerical,%
]{revtex4-2}
\usepackage[english]{babel}
\usepackage{graphicx}
\usepackage{dcolumn}
\usepackage{bm}
\usepackage[mathlines]{lineno}

\usepackage{multirow}

\begin{document}


\title{Characterising the role of human behaviour in the effectiveness of contact-tracing applications}

\author{Ariadna Fosch}
 \email{arifosch@gmail.com}
 \affiliation{CENTAI Institute Turin Italy}
 \affiliation{Institute for Biocomputation and Physics of Complex Systems \\ (BIFI) University of Zaragoza Spain}
 \affiliation{Department of Theoretical Physics University of Zaragoza Spain}
\author{Alberto Aleta}%
 \affiliation{Institute for Biocomputation and Physics of Complex Systems \\ (BIFI) University of Zaragoza Spain}
 \affiliation{Department of Theoretical Physics University of Zaragoza Spain}

\author{Yamir Moreno}
 \affiliation{Institute for Biocomputation and Physics of Complex Systems \\ (BIFI) University of Zaragoza Spain}
 \affiliation{Department of Theoretical Physics University of Zaragoza Spain}


\begin{abstract}
Albeit numerous countries relied on contact-tracing (CT) applications as an epidemic control measure against the COVID-19 pandemic, the debate around their effectiveness is still open. Most studies indicate that very high levels of adoption are required to stop disease progression, placing the main interest of policymakers in promoting app adherence. However, other factors of human behaviour, like delays in adherence or heterogeneous compliance, are often disregarded. To characterise the impact of human behaviour on the effectiveness of CT apps we propose a multilayer network model reflecting the co-evolution of an epidemic outbreak and the app adoption dynamics over a synthetic population generated from survey data. The model was initialised to produce epidemic outbreaks resembling the first wave of the COVID-19 pandemic and was used to explore the impact of different changes in behavioural features in peak incidence and maximal prevalence. The results corroborate the relevance of the number of users for the effectiveness of CT apps but also highlight the need for early adoption and, at least, moderate levels of compliance, which are factors often not considered by most policymakers. The insight obtained was used to identify a bottleneck in the implementation of several apps, such as the Spanish CT app, where we hypothesise that a simplification of the reporting system could result in increased effectiveness through a rise in the levels of compliance.
\end{abstract}

\maketitle
\section{Introduction}
Contact tracing (CT) is one of the most effective epidemic control strategies, as it allows cutting the disease transmission chains by isolating potentially infected individuals before they can further spread the pathogen~\cite{SARS}. However, during the early days of the COVID-19 pandemic, this classical strategy was hardly effective due to the long turnaround of tests, and the presence of presymptomatic ineffectiveness and mild symptomatology in a large proportion of cases. As a consequence, many countries promoted the use of novel digital contact tracing (DCT) strategies, based on the use of smartphone apps. These apps, also known as CT apps, rely on several technologies to register interactions between individuals and warn those that recently had contact with someone who turns out to be infected. This way, these individuals could quarantine themselves preventively, much sooner than with classical contact tracing~\cite{Ferretti2020May}.

These apparent benefits produced an explosive growth of CT apps even though there was no empirical evidence of the effectiveness of these tools~\cite{braithwaite}. In less than a year, researchers from MIT identified at least 80 different CT apps deployed over 50 countries~\cite{ONeill2020Oct}. Concurrently, many modelling studies tried to understand whether these apps could be effective, and it was generally accepted that the adoption rate should be around 60\% of the population for pandemic mitigation~\cite{Ferretti2020May, vittoria, Pandit2022Jul}. Nonetheless, current empirical evidence signals that in practice most apps were not as effective as expected and that many models were too optimistic~\cite{Burdinski2022Dec}.

An important issue that is usually neglected is the fact that, in many cases, the user has to manually go into the app and alert her list of contacts, which greatly hinders the efficacy of the procedure even for large levels of adoption. For instance, authorities in New South Wales, Australia, analyzed the outcome of using their app versus conventional contact tracing during an outbreak of 619 cases. First, only 137 of them had the app, which corresponds to an adoption rate of 33\%. Then, only 32 warned at least one contact using it. In total, there were 79 contacts notified, or roughly 3 per index case. In contrast, conventional contact tracing revealed 25,300 close contacts, 40 per index case. The app only detected 17 ($< 0.1\%$) close contacts not identified through conventional tracing. The Australian health workers considered that the app was not useful for them and that actually increased their burden~\cite{Vogt2022Mar}.

Similar conclusions were reached in other countries. In Finland, out of 4,557 PCR-positive cases, only 541 warned their contacts (12\%). Besides, most people that received the warning had already been alerted through traditional contact tracing, since the procedure to notify contacts with the app was rather slow. In total, only 8 (0.3\%) people reported having changed their behaviour due to the app notification~\cite{Rannikko2022Jun}. In Switzerland, with an adoption rate of 26\%, only between 20 to 40\% of the users triggered an exposure notification upon receiving a positive test~\cite{Daniore2022Nov}. In Belgium, the adoption rate was 28\%, and 43\% of the users employed it to notify their contacts~\cite{Proesmans2022Dec}. In France, adoption was around 25\%~\cite{Schultz2022May}. 

Overall, adoption was relatively low, and in any case far from the targeted value of 60\%. The low amount of people using the app to alert their contacts - either due to technical difficulties with the notification system or due to their unwillingness to do so - further diminished their effectiveness. However, a much more positive conclusion was reached in the UK. In that case, adoption was close to 29\%. During an outbreak of over half a million individuals, 1.7 million notifications were sent to potential contacts, roughly 3 per case. With an estimated secondary attack rate of 6\%, and assuming that 65\% of individuals adhere to quarantine, it was estimated that between 4,200 to 8,700 deaths were averted by the app, claiming that it was highly successful~\cite{wymant}.

As we can see, both the adoption and the number of contacts alerted per index case were very similar to the ones found in other countries. Thus, the main difference is in the size of the outbreak rather than in the app characteristics. While the outbreak in Australia was relatively small and conventional contact tracing was extraordinarily effective, identifying 40 close contacts per index case, manual contact tracing in the UK led to the identification of 2 close contacts, on average~\cite{Poletto2022Mar}. Similarly, this number was 1.4 in France and less than one in parts of the United States~\cite{Lewis2020Dec}. Hence, when incidence is low, CT apps may only provide a marginal benefit.

Another important aspect that has been observed empirically but that is often disregarded in CT app models is the presence of heterogeneities in the app adoption process and its use~\cite{Munzert2021Feb,OConnell2021Jun,guazzini}. App adoption is a dynamic process, where individuals can install/uninstall the app at will, changing their decision as the epidemic evolves. Therefore, it is crucial to create more complex models that consider dynamic adherence and heterogeneities in compliance to understand the actual role that CT apps can play in controlling future epidemics.

To this end, we propose a hybrid model that combines epidemic and human behaviour dynamics to describe the co-evolution of a disease outbreak and the CT app adoption process over a single population. Its implementation relies on a multi-layer network structure, as it allows us to represent the separate connectivity patterns in which each dynamic evolves (in-person contacts and Bluetooth interactions) while still establishing a functional coupling between them. This novel approach expands prior CT app models by considering app adoption as a dynamic decision-making process and also by including heterogeneity in compliance with the reporting system. 

\section{Methodology}

\subsection{Network structure}\label{network}
To test the effectiveness of CT apps in mitigating epidemic spreading, we relied on two dynamical models (epidemic spreading and CT app adoption) co-evolving over a single population of 10,000 individuals. The population structure was represented through a two-layer multilayer network, where one layer represents the in-person contact network in which the epidemic evolves and the other one the mobile-phone interactions registered by the CT app, which are only a proxy for in-person contacts. As both dynamics are occurring over the same individuals, a bijective (one-to-one) mapping was introduced to couple together the nodes in both layers. See figure~\ref{overview}A for a schematic representation of the model.

The degree distribution of the in-person contact network was defined to follow a negative binomial distribution fit with survey data from the Italian population~\cite{mossong, mistry}. To fit the distribution, we followed the approach described in Ref~\cite{lu}, which estimates the average degree, $\langle k \rangle$ of the distribution from the age-mixing matrices in Ref~\cite{mistry} ($\mu = 11.92$), and the dispersion parameter ($r$) from the age-aggregated data from the POLYMOD study ($r=2.426$)~\cite{mossong}.

We created the CT app interaction network by expanding the in-person contact network with random edges. These spurious links reflect interactions meeting the inclusion criteria of the CT app (location and duration) but do not entail a risk of infection, because of other protective measures or physical barriers between the users. For instance, in Australia, up to 61\% of the contacts identified by the app were workers in adjacent rooms, customers in neighbouring restaurants or even people waiting in separate cars at COVID-19 drive-through testing clinics~\cite{Vogt2022Mar}. Thus, we duplicated the number of contacts associated with random encounters in this network. Given that random contacts encompass approximately 25\% of the 11.92 daily interactions~\cite{mistry}, this yields an average degree in the CT app network of $\langle k \rangle_{\text{App}}=14.81$.

To evaluate the influence of the population structure in the results obtained we repeated the analysis for populations with two artificial degree distributions (Erd\H{o}s-Rényi random graph~\cite{erdos} and a Scale-Free network~\cite{barabasi}) and a very similar $\langle k \rangle$. More details about the network definition and the fitting process are available in Supplementary Materials (SM) Sec.~I.A.

\subsection{Epidemic model} \label{epidemic}
We described the natural history of SARS-CoV-2 using the model shown in figure~\ref{overview}C. It is based on the Susceptible-Exposed-Presymptomatic-Infected-Removed (SEPIR) model, which is an extension of the SEIR model that accounts also for presymptomatic infection~\cite{Anderson1992Aug}. The SEPIR model assumes susceptible individuals ($S$) can be infected by interacting with neighbouring individuals in any of the infectious states, pre-symptomatic ($P$) or infected ($I$). After contagion, $S$ individuals transition towards the exposed state ($E$), in which they are already infected but not yet infectious. Once they become infectious, despite still not showing symptoms, they switch to the pre-symptomatic infected state ($P$). As the disease progresses, symptoms appear and $P$ individuals enter the symptomatic infectious state ($I$), in which they will remain until they either overcome the disease or die, thus transitioning to the removed state ($R$).

We modified the SEPIR model by introducing quarantined equivalents to the $S$, $E$, $P$, and $I$ states ($S_q$, $E_q$, $P_q$, and $I_q$). In them, disease follows the same progression as in their non-quarantined counterpart, but individuals in those states are not contagious. Transitioning from a free state to its quarantined analog occurs mainly through the CT app's warning system. All individuals that receive a message from the app are forced to enter a 10-day preventive quarantine regardless of their current state. Additionally, a fraction of the daily new symptomatic individuals ($P \rightarrow I$ transitions) are detected in healthcare testing and quarantined one day after developing symptoms. By limiting healthcare detection to symptomatic cases on their first day of symptoms we reflect the large fraction of individuals with mild or asymptomatic COVID-19 manifestations, who would never be detected without additional control strategies. Contrary to the preventive quarantines, confirmed positive cases ($I_q$) remain under quarantine until their complete recovery. 

The transition rates across compartments were defined based on epidemiological parameters estimated during the 1st wave of the COVID-19 pandemic. Prior research has identified the median incubation period of SARS-CoV-2 to 5.1 days (95\% CI, 4.5 to 5.8 days)~\cite{lauer} and the generation time to 7 days approximately. Thus, we assumed the transition rates between $E\rightarrow P$ and $P\rightarrow I$ to be $\epsilon = 1/3$ and $\rho=1/2$ respectively, resulting in an average incubation period of $5$ days. To reflect the 7-day generation time, the transition rate from $I \rightarrow R$ was assumed to be $\mu=1/2$. Finally, we set the transmissibility parameter ($\beta$) according to the basic reproductive number ($R_0$) estimated in European countries during the 1st wave of the COVID-19 pandemic ($2<R_0<3$)~\cite{hilton}. The relationship between $\beta$ and $R_0$ is given by

\begin{equation}
    R_0= \beta \left(\frac{1}{\rho}+{\frac{1}{\mu}} \right) \lambda_{\mathbb{R}}(A_{ij})
    \label{R_0_Eq}
\end{equation}
where $\beta$, $\rho$, and $\mu$ represent the epidemic parameters and $\lambda_{\mathbb{R}}$ is the real component of the largest eigenvalue of the network's adjacency matrix~\cite{diekmann}. Hence, for a given $\beta$, the actual value of $R_0$ will depend on the network under consideration. We calibrated the model using the network built with a negative binomial degree distribution, since it is the most realistic one, to obtain $R_0=3$. This yields $\beta = 0.045$, see figure~\ref{meanR0}B. The same $\beta$ was used for the other two networks so that any differences will be directly related to the network configuration and not to different epidemiological parameters. The resulting outbreaks differ in the $R_0$ value but show a similar prevalence for the range of $\beta$ under consideration, Figure~\ref{meanR0}A.

To define the transition rates for healthcare detection and the duration of preventive quarantines we also relied on empirical data. Seroprevalence studies have shown that during the first wave only 10\% of the cases were detected, increasing to 60-70\% in subsequent waves~\cite{Pollan2020prevalence}. Thus, we assumed that half of the new daily symptomatic were detected on the first day of infection ($\delta = 0.5$) and that the duration of the quarantines is 10 days, as suggested in the COVID-19 quarantine protocol followed by the Spanish government during most of the pandemic~\cite{protocolo}.

\subsection{Contact-tracing app model}\label{sec:CTapp}
\begin{figure}[!ht]
\begin{center}
\includegraphics[width=\columnwidth]{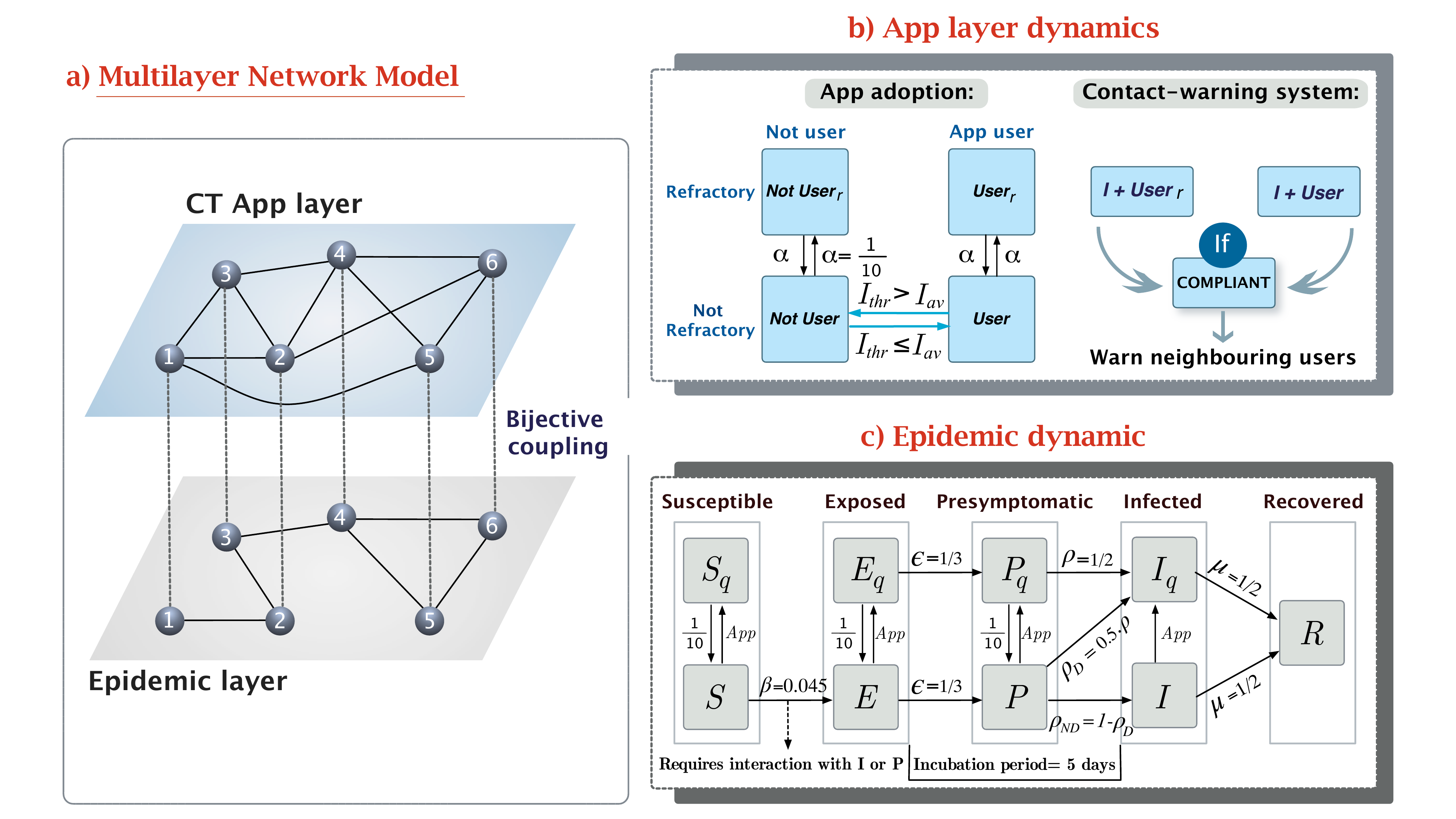}
\end{center}
\caption[Overview of the epidemic-CT app multilayer network model]{ Overview of the epidemic-CT app multilayer network model. (\textbf{A}) Schema of the multilayer network structure and the coupling between layers. (\textbf{B}) Compartmental model of the dynamic in the CT app layer. It contains two separate dynamics, the app adoption dynamic, and the app effect. \textbf{C} Compartmental model for the epidemic layer. It is based on a modified SEPIR epidemic model~\cite{Anderson1992Aug}, with the addition of a presymptomatic state ($P$) and quarantined versions of the $S$, $E$, $P$ and $I$ states. The transition rates have been defined to produce outbreaks reassembling the 1st wave of the COVID-19 pandemic.}\label{overview}
\end{figure}

\begin{figure}[!ht]
\centering
\includegraphics[width=0.75\columnwidth]{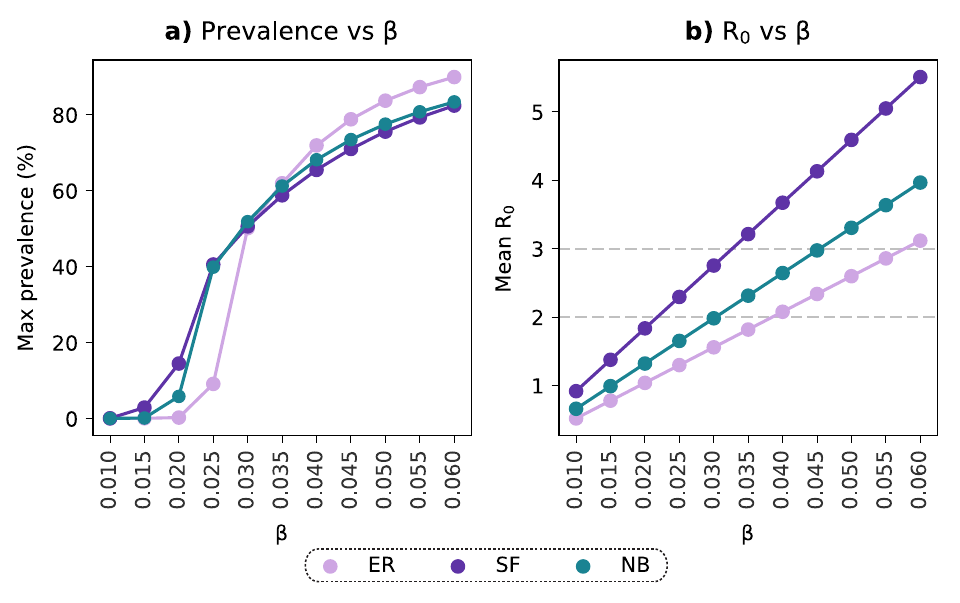}\vspace{-0.2cm}
\caption[Plots for model initialisation]{Selection of the transmissibility parameter on the basis of $R_0$. \textbf{(A)} Maximal prevalence for simulations with ranging values of $\beta$. \textbf{(B)} Basic reproductive number estimated for simulations with varying $\beta$. Light-violet trends represent the results for the Erd\H{o}s-Rényi connectivity matrix (ER), purple trends reflect estimations for the Scale-Free (SF) population and the blue points reflect the results for the Negative Binomial distribution (NB) fitted to data from contact patterns surveys.}
\label{meanR0}
\end{figure}

As shown in figure~\ref{overview}B, the dynamical model in the CT app layer reflects two different processes: the app adoption dynamic and the warning system.

The first dynamics describes the decision-making process of each individual to download or remove the CT app depending on their reluctance level and the pressure of external factors. Prior research has described binary decision-making in rational individuals (like adherence to riots or trends) using threshold dynamics, where individuals adopt the behaviour if the adoption pressure overcomes their reluctancy threshold and do not adopt it otherwise~\cite{granovetter}. Our model assumes that disease progression acts as a positive pressure towards app adoption, as it has been observed that fear of infection will push more reluctant individuals to overcome their concerns and download the app. Thus, the reluctancy threshold for each individual was defined as the minimal level of infection (7-day incidence/100,000 inh.) triggering their adoption of the app. 
Individuals will only use the CT app while they feel at risk (incidence level above their reluctancy threshold), and they will uninstall it once the incidence returns below threshold. Generally, this decision is not made on a daily basis, even if the environmental conditions have changed. Thus, we introduced a refractory period after each choice equal to the duration of the preventive quarantine, 10 days. 

We introduced heterogeneity between individuals by sampling their reluctancy threshold from a Poisson distribution with a pre-defined average reluctancy level. This distribution was modified in both extremes to include individuals with extreme responses: 

\begin{itemize}
    \item Non-adopters: Individuals who would never adopt the app, either because they are too concerned about their privacy, or they do not have access to a compatible smartphone. In the case of European countries, prior studies have shown that at least 30\% of the population may be unable to acquire CT apps~\cite{vittoria}. We use this value to set the upper bound in the maximal number of adopters to 70\%.
    
    \item  Early adopters: Pioneer individuals without reluctance towards app adoption. They will download the app right after its implementation and only uninstall it after the complete extinction of the epidemic outbreak ($I=0\ \text{cases/100,000 inh.}$). We considered only 1\% of the population to be early adopters.
\end{itemize}

The population’s average reluctancy threshold ($I_{\text{thr}}$) can be interpreted as the cultural differences between different populations, in terms of their willingness to adopt the CT app. Populations with a high average threshold require more time to reach the level of infection triggering the generalised adoption of the apps, while low threshold populations will adopt the CT app more easily. 

The second dynamics implemented in the CT app layer is the reporting system. Only compliant app users that test positive for SARS-CoV-2 (individuals transitioning to the $I_q$ state) will report their infection in the CT app and activate the contact warning system. This will cause all their neighbours in the app layer who are also active users to enter a 10-day preventive quarantine (as described in Sec.~\ref{epidemic}). We assumed that compliance is only related to reporting, thus, individuals who are labeled as non-compliant will still quarantine themselves since they willingly downloaded the app. Individuals who do not report their status nor follow the app recommendations are those that do not have the app installed.

A list of the parameters used for the epidemic-CT app model is available in SM Sec.~I.B 

\subsection{Simulations}
We assessed the impact of the CT app intervention by measuring the difference in peak incidence and total prevalence between a simulation with the CT app and a baseline scenario without it. Both simulations were initialized with the same initial conditions and iterated for 500 time steps (500 days), enough time to observe a complete dynamic in both layers. To reduce the effects of stochastic noise, we repeated each simulation 1,000 times starting from different initial conditions.

Before estimating the effectiveness of the CT app, we removed all repetitions not resulting in an effective outbreak ($\text{max}(I)<1\%$) and aligned the surviving ones following Ref.~\cite{kiss} (see SM Sec.~I.C for a description of the alignment). The effectiveness of the CT app was measured over the average response of the surviving repetitions in each simulation and it was estimated by using the relative reduction in peak incidence ($\Delta_i$) and maximal prevalence ($\Delta_p$). For both cases $\Delta$ was defined as

\begin{equation}
\label{delta}
\Delta = 1-\frac{\text{max}(I_{\text{ CT app}})}{\text{max}(I_{\text{ Baseline}})}
\end{equation}
where $I_{\text{ CT app}}$ is the value of the metric (peak incidence or maximal prevalence) in the scenario with an active CT app, and $I_{\text{Baseline}}$ is the same metric for the scenario without it.

The relative peak incidence reduction is associated with a flattening of the epidemic curve, indicating the potential of the CT app to reduce the speed of the epidemic propagation and the pressure on the healthcare system. Meanwhile, maximal prevalence reduction is an indicator of the overall impact of the CT app in reducing the total number of infections, providing a more global perspective of the effect of the app. 

\section{Results}\label{results}
\subsection{Scenario description}
We evaluated the influence of three factors of human behaviour in the effectiveness of CT apps: the maximal percentage of adoption, the average reluctancy towards app adoption and the fraction of compliant users. To this end, three hypothetical scenarios were implemented: the voluntary adoption scenario (where 100\% compliance is assumed), the imposed adoption one (with zero reluctance towards app adoption) and the ``adherence \& compliance" scenario (only constraining the maximal level of adoption). These scenarios can also be interpreted according to the characteristics of the app adoption campaigns followed by different countries during the COVID-19 pandemic. In  European countries, most countries relied on voluntary adoption campaigns, where adherence is not compulsory and it only depends on the population's willingness to download the app (its average reluctance threshold). In this scenario, complete compliance is assumed under the hypothesis that individuals who voluntarily decide to adhere to the strategy will also be more likely to comply with it. Contrarily, the imposed adoption strategy assumes that individuals are forced to download the app from the start of the epidemic outbreak ($I_{\text{thr}}=0\ \text{cases/100,000 inh.}$) but no control is exhorted over their use.

\begin{figure}[!ht]
\begin{center}
\includegraphics[width=\columnwidth]{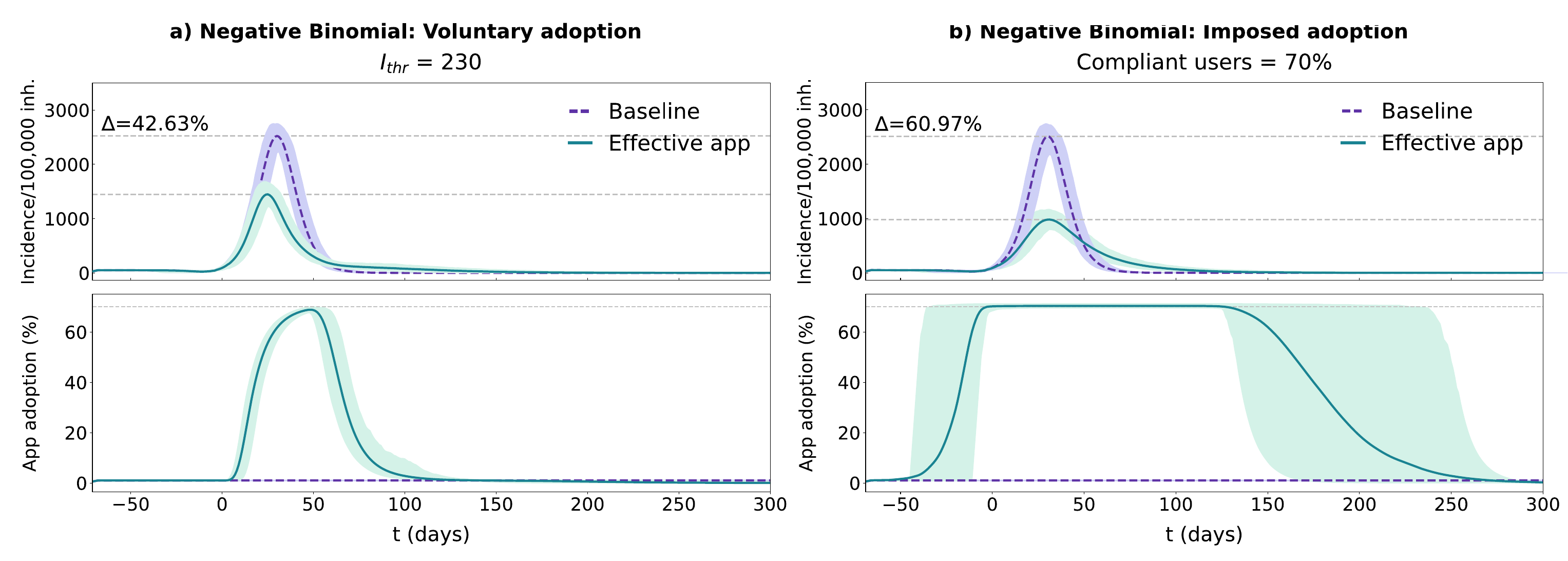}
\end{center}
\caption{Temporal evolution for the epidemic and app adoption dynamics in the voluntary adoption (panel \textbf{A}) and imposed adoption (panel \textbf{B}) scenarios in a population with a realistic distribution (Negative Binomial fit with survey data). In each panel two conditions are tested, purple traces reflect the baseline condition (no app adoption) and the green ones reflect the effective CT app situation. The results for both conditions are represented with the average and 95\% CI across 1000 repetitions. Epidemic progression is reported using the 7-day incidence/100,000 inh. (more details available in SM Sec.~I.D), while CT app adoption is reported with the percentage of the population using the CT app. (\textbf{A}) Simulation for the voluntary adoption scenario with an $I_{\text{thr}}= 230\ \text{cases/100,000 inh.}$. (\textbf{B}) Simulation for the imposed adoption scenario with a fraction of compliant users $=70\%$.}\label{scenarios}
\end{figure}

The differences between these two approaches are more evident when comparing their evolution over the same baseline epidemic outbreak. In the voluntary adoption scenario (figure~\ref{scenarios}A), app adoption only starts rapidly growing after the epidemic outbreak has reached the population's average reluctancy threshold ($I_{\text{thr}}=230\ \text{cases/100,000 inh.}$ in this case). As the epidemic progresses, the number of users continues to increase, reducing its adoption rate as the epidemic starts its decline. When the incidence level decreases below the threshold, the number of users rapidly declines, until the complete removal of the app around $t=120$. In this scenario, the average reluctancy threshold controls the start of the adoption process and the time of removal of the app, conditioning the total duration of the control strategy. Lower reluctancy produces wider windows of app adoption, resulting in higher detection rates and increased app effectiveness. This can be observed in more detail in Supplementary figure~3A. 

In the imposed adoption scenario the app adoption dynamic differs (see figure~\ref{scenarios}B). Since the average reluctancy threshold is defined to $I_{\text{thr}}=0\ \text{cases/100,000 inh.}$, app adoption rapidly grows at the start of the simulation, reaching the maximal level of adoption ($70\%$) before the epidemic's exponential growth phase ($t=-4$). Maximal adoption is maintained for the rest of the epidemic outbreak and the app removal cascade will only start after the complete extinction of the disease. In this scenario, the maximal number of adopters is maintained for the whole duration of the epidemic, regardless of its size (see Supplementary figure~3B). Thus, compliance affects the effectiveness of the strategy through a reduction in the performance of the reporting system, not through a change in the duration of the adoption window.

Finally, the ``adherence \& compliance” scenario relaxes the assumptions of complete compliance and no reluctance towards app adoption assumed respectively, in the voluntary and imposed adoption scenarios. It represents a more realistic scenario, where the only constraint introduced over the human behaviour parameters is on the upper bound of the number of adopters (set to 70\% of the total population). As mentioned in Sec.~\ref{sec:CTapp}, the number of app users is intrinsically upper-bounded by the proportion of individuals who do not have an appropriate device (30\% of the population).

\subsection{Exploratory analysis of the human behaviour parameters }
Figure~\ref{heatmaps} shows the results for the voluntary and imposed adoption scenarios and figure~\ref{adcop} shows the results for the  ``adherence \& compliance" simulation. The effectiveness of the apps is measured in terms of peak incidence reduction ($\Delta_i$) and prevalence reduction ($\Delta_p$). To facilitate the interpretation of the results we defined qualitative performance patterns depending on the incidence and prevalence reduction induced by the app: no effectiveness ($\Delta<5\%$), low effectiveness ($5\%<\Delta<10\%$), moderate effectiveness ($10\%<\Delta<20\%$) and high effectiveness ($\Delta>20\%$).

The analysis for the voluntary adoption scenario (figure~\ref{heatmaps}A and C) revealed that CT apps are only effective ($\Delta>5\%$) when low reluctancy thresholds and high levels of adoption are present. Strategies with late adoption ($I_{\text{thr}} \sim 1300\ \text{cases/100,000 inh.}$) are only minimally effective in peak-incidence reduction when they are adopted by a very large fraction of users ($>50\%$). However, late adoption is less relevant for prevalence reduction, where the same combination of parameters resulted in a moderate prevalence reduction ($\Delta_p>10\%$). Interestingly, to obtain a high effectiveness strategy for peak incidence reduction ($\Delta_i>20\%$) it is not enough to have more than 50\% of adopters, the population's average reluctancy threshold must also be very low ($I_{\text{thr}} < 400\ \text{cases/100,000 inh.}$). 

In the case of the imposed adoption scenario, we observed that CT apps with high levels of adoption and moderate levels of compliance are necessary to obtain effective strategies ($\Delta > 5\%$). Apps with less than 10\% of compliance are mostly ineffective, even when considering high levels of adoption. The relevance of compliance for app effectiveness is even more apparent when aiming towards a high-effectiveness strategy ($\Delta > 20\%$). Even when assuming a high level of adoption ($50\%$), it is only possible to obtain a high peak incidence reduction if more than 30\% of compliance is guaranteed. 

The results for the ``adherence \& compliance" scenario (figure~\ref{adcop}) confirm the trends observed in the previous strategies, evidencing that moderate levels of compliance ($>15\%$) and low reluctancy thresholds ($I_{\text{thr}}<800\ \text{cases/100,000 inh.}$) are required to obtain relevant peak incidence reductions, and that even strategies with a high reluctancy threshold can result in moderate prevalence reductions if almost everyone complies with them.

The analysis for the artificial networks (Erd\H{o}s-Rényi and Scale-Free) populations can be found in Supplementary figures 4-7 and they support the conclusions extracted from the realistic population. 

\section{Discussion}
The first models of the effectiveness of CT apps for pandemic control were fairly optimistic. They modelled app adoption through a static point of view, assuming a constant amount of users for the whole duration of an outbreak. In reality, app adoption is a dynamic process, where individuals can decide to download and remove the app at will as the epidemic progresses. Our study proposes a novel approach for representing CT apps, where app adoption is modelled as a threshold dynamic depending on epidemic progression and heterogeneities in the reluctance threshold. Besides, we also include the effect of the level of compliance with the reporting system. Note that compliance can represent both the own willingness of an individual to report her status or the inability to do so due to technical issues.

Using this model we explored the interplay between app adoption and epidemic progression and characterized how different human behavioural heterogeneities alter the performance of CT apps. This was achieved by simulating three separate scenarios: the voluntary adoption (assuming complete compliance) the imposed adoption (assuming no reluctancy towards app adoption) and the ``adherence \& compliance'' scenario, where the only constraint is the maximum number of people who can download the app (70\% of the total population).

\begin{figure}[!ht]
\begin{center}
\includegraphics[width=\columnwidth]{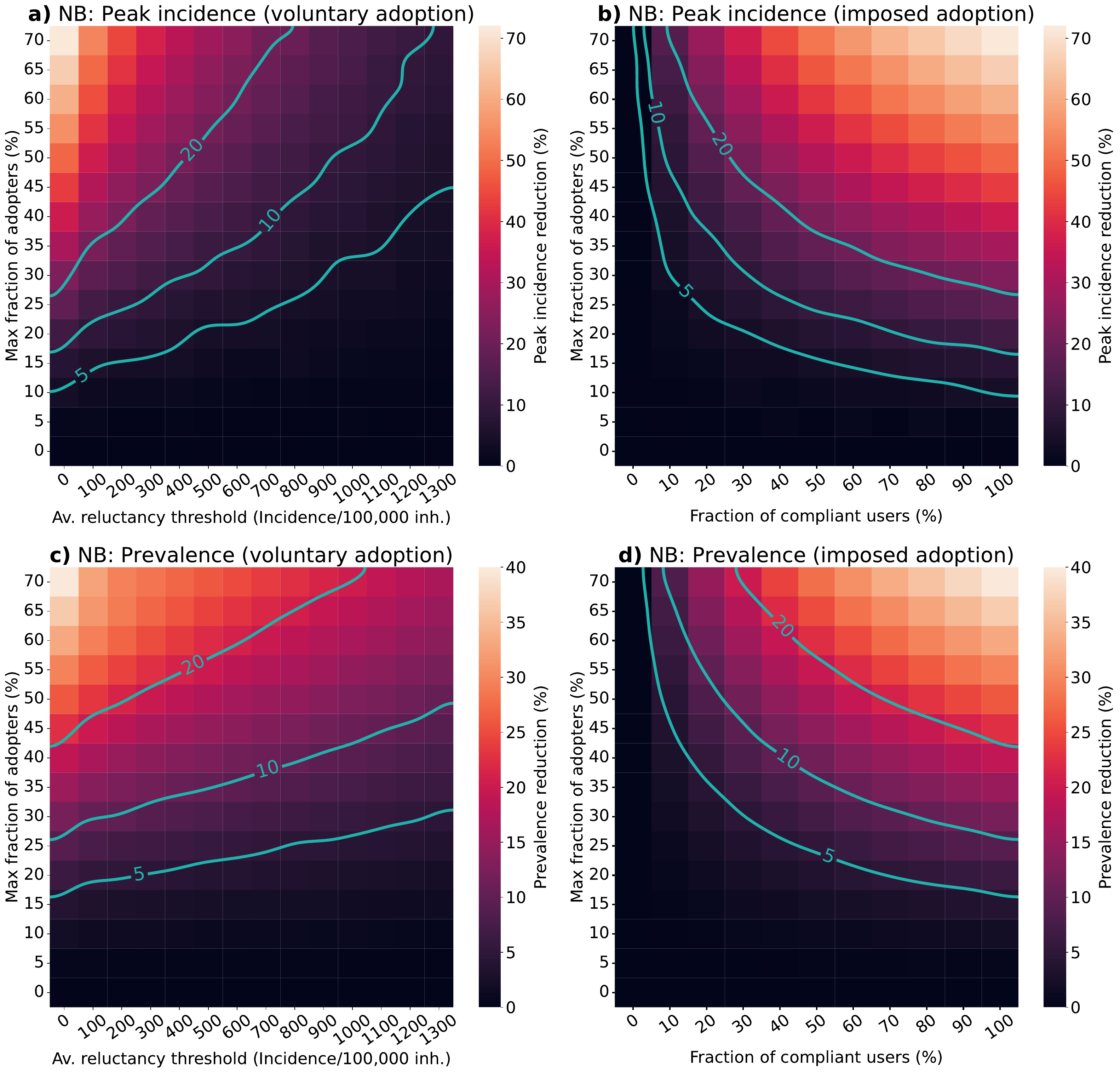}
\end{center}
\caption{Impact of different factors of human behaviour in the effectiveness of CT apps in a population with a Negative binomial distribution (NB). For the voluntary adoption scenario (panels \textbf{A} and \textbf{C}), the parameters explored are the average reluctance threshold and the maximal fraction of adopters. Meanwhile, in the imposed adoption scenario (panels \textbf{B} and \textbf{D}), changes in the fraction of cooperative users and the maximal fraction of adopters are explored instead. The colour scale reflects the average reduction produced by the CT app ($\Delta$) in the peak incidence (top panels) or maximal prevalence (lower panels). The isoclines indicate the combinations of parameters resulting in $\Delta=5\%$, $\Delta=10\%$ and $\Delta=20\%$.}\label{heatmaps}
\end{figure}

\begin{figure}[!ht]
\begin{center}
\includegraphics[width=\columnwidth]{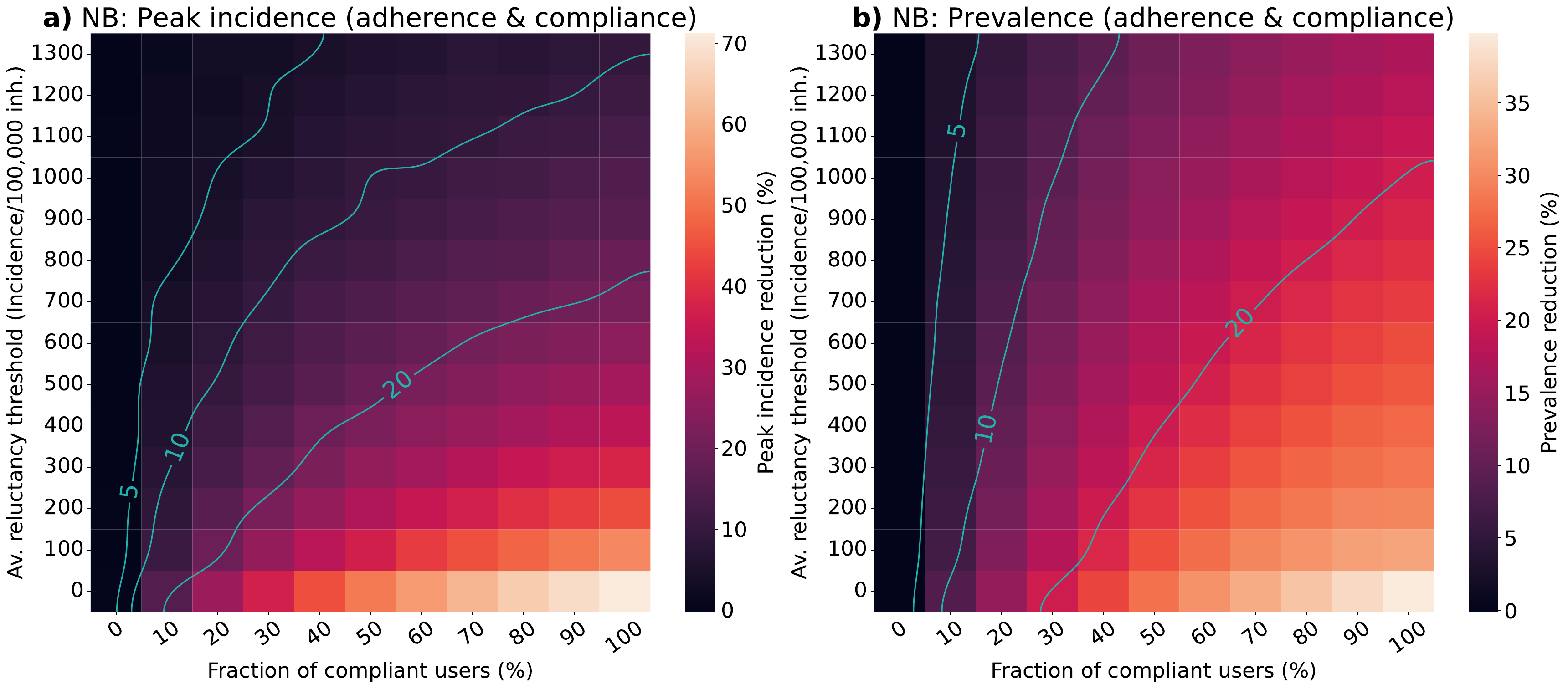}
\end{center}

\caption{Impact of the population's average reluctance threshold and the level of compliance in the effectiveness of CT apps for a population with a Negative binomial distribution (NB). We assumed a maximal fraction of adopters of 70\% for the ``adherence \& compliance" scenario. The colour scale reflects the average reduction produced by the CT app ($\Delta$) in the peak incidence (panel \textbf{A}) or maximal prevalence (panel \textbf{B}). The isoclines indicate the combinations of parameters resulting in $\Delta=5\%$, $\Delta=10\%$ and $\Delta=20\%$.}\label{adcop}
\end{figure}

The exploratory analysis for the voluntary and imposed adoption scenarios confirmed the necessity of high levels of adoption to obtain effective CT apps (figure~\ref{heatmaps}). Apps adopted by less than 10-15\% of the total population always result in ineffective strategies regardless of the time of adoption and the level of compliance. This lower bound of adoption corresponds to approximately 21\% of all smartphone users, a value very close to the minimal penetration needed identified in prior studies ($20\%$ in Ref~\cite{vittoria}).

Even with more than 50\% of maximal penetration, high effectiveness strategies are only obtained when the reluctancy threshold is low ($< 400$ cases/100,000 inh.) in the voluntary adoption scenario, or compliance is moderate or high ($>20\%$ compliant users). Note that incidences of the order of 1,000 cases/100,000 inh. were common in Europe through 2020-2022 and still app adoption was close to 20-30\%. Similarly, even though people voluntarily downloaded the app, compliance was in the range of 20-40\%, and thus one would expect lower values if adoption is imposed. As such, high effectiveness may not be reachable in empirical settings. Nonetheless, for large enough outbreaks - such as the one observed in the UK - even low reductions of prevalence may result in a noticeable decrease in hospital burden. 

The results shown in figures \ref{heatmaps}B and \ref{heatmaps}D also highlight the importance of increasing user compliance. Limited research has addressed the importance of reporting compliance for DCT. The study from Ref.~\cite{vittoria} for France explored the impact of heterogeneities in quarantine compliance, but still assumed perfect compliance with reporting. An interesting study about compliance with infection reporting is Ref~\cite{davi}, which explored how adherence and compliance can play a relevant role in a traditional contact-tracing strategy based on self-reporting. In their scenario with low self-reporting (around 11\%), scalability did not have a significant impact on the overall effectiveness. This was also observed for our model, where increasing the percentage of adoption did not result in major improvements in app effectiveness for compliance levels below $20\%$. Adherence only starts playing a major role in increasing app effectiveness if moderate levels of compliance are ensured. 

The findings from this study may have implications for policymakers seeking to use DCT strategies for future epidemic outbreaks. While current promotion efforts mainly aim to increase app usage, there is little emphasis on the importance of early adoption and proper usage. We hypothesize that promoting the importance of compliance and early adoption could lead to an increase in CT app effectiveness against future pandemics.  Besides, our findings could also be valuable for researchers proposing novel DCT implementations. Evaluating the sensitivity to reporting compliance before the app's implementation can aid in the creation of more robust systems. 

Compliance may also be related to the very implementation of the app. To exemplify this point, we analyzed the performance of RadarCOVID, the Spanish CT app~\cite{radar}. According to their statistics, after 40 weeks of implementation, the app had a 19\% of penetration and only 6.7\% of the users were able to report their infection~\cite{radar}. Our model shows that this combination of parameters results in a completely ineffective strategy both in terms of peak incidence and prevalence reduction (figure~\ref{heatmaps}B and D). The low levels of compliance in the Spanish app may have resulted from its complex reporting system. Users needed to obtain a verification code from the regional healthcare authorities and enter it into the CT app to report their infection. However, the codes were generated by the central Health Ministry, which had to communicate with regional authorities~\cite{radar_tecnico}. Due to this interaction, there were significant delays in reporting infections, and many users did not receive their verification code even after requesting it. 

We hypothesize that apps that follow a similar approach could benefit from a more straightforward reporting strategy, where the healthcare authorities are responsible for activating the positive status of an individual once the user has provided their consent, removing the responsibility from the final user~\cite{melder}. This would ensure almost perfect compliance at the expense of maybe increasing privacy concerns towards app adoption. Hence, there is no perfect solution and more research should be devoted to understanding how DCT can actually be implemented in practice, taking into account the complexities of human behaviour. 

\subsection{Limitations and Future research}
One of the main considerations of the study is the definition of the CT app model. App adoption was assumed to follow threshold dynamics only dependent on the prior incidence and the reluctance level of each individual. This follows the hypothesis that media reporting about disease progression can act as a driving force to encourage more reluctant individuals to download the app for their protection. There is still not a clear consensus about the main driving factor for app adoption. Ref~\cite{guillon} suggests CT app adoption in France was not severely influenced by the risk perception of their users. They propose that concerns about data protection and misinformation played a more relevant role in shaping app adoption than the perception of being at risk. However, Ref~\cite{nguyen}, a study based on technology acceptance models (TAM), states that health risk perception has a positive effect on CT app adoption. Future research could aim towards developing an adoption model combining health risk perception and peer pressure dynamics.

Another limitation of our study involves the app reporting process. We assumed immediate and perfect reporting and testing, with infectees warning their contacts the same day of the diagnosis and the quarantine starting the day after the alert. Prior research has identified that delays in the reporting process, poor compliance with the preventive quarantine and testing unavailability can drastically diminish the effectiveness of CT app strategies~\cite{Ferretti2020May,kretzschmar, almagor}. Furthermore, empirical evidence shows that even if compliance may be of the order of 40-70\% of the users, only 10-50\% of the ones that receive the alert contacted the authorities~\cite{Burdinski2022Dec}. In our case, we assumed that the latter was 100\%. Thus, our results should be interpreted as an upper bound of the performance that can be achieved in realistic settings. 

Future research could also be directed towards creating more realistic population structures. It is known that age heterogeneities can play a major role in the distribution of reluctance towards app adoption~\cite{vittoria}, thus it would be possible to modify the distribution of reluctance to better describe the patterns observed in realistic populations. Prior studies have reported that a high CT app coverage amongst adults plays a central role to prevent transmission to the elderly, who have less accessibility to smartphones~\cite{vittoria}. By including age heterogeneities in our model it would be possible to study the impact of delayed adoption and low compliance in the protection of the elderly. This question could provide more insight into the realistic effectiveness of DCT and could aid policy-makers in better controlling future outbreaks.

\subsection{Conclusion}
Overall, this study presents a novel approach to represent the co-evolution of epidemic progression and CT app adoption from a dynamic point of view. This approach allowed us to better characterize CT apps by including human behavioural heterogeneities like the individual's reluctance towards app adoption or different levels of compliance. The analysis evidenced the importance of examining beyond the maximal level of adoption for obtaining effective CT apps. Early adherence and moderate levels of compliance are paramount for effective CT apps, as could be observed in the case of the Spanish CT app, where really low levels of compliance make the app almost ineffective. 

Furthermore, even if the model included many new components related to human behaviour, in many aspects it was still too optimistic. For instance, quarantine compliance was assumed to be perfect and case detection or information propagation was also much higher than the ones observed empirically. As such, the results must be interpreted as an upper bound on the performance of these apps. Given the low impact that we observed for values in the empirical range, this implies that CT apps can hardly be used as the only protective measure during an outbreak. Instead, they can be used to complement other mitigation strategies, such as classical contact tracing. Lastly, more research should be directed toward integrating the complexities of human behaviour in epidemic processes, and the implementation of these apps should be revisited in light of the evidence obtained for future pandemics.

\section*{Conflict of Interest Statement}
The authors declare that the research was conducted in the absence of any commercial or financial relationships that could be construed as a potential conflict of interest.

\section*{Author Contributions}
\textbf{Ariadna Fosch:} Software, Formal analysis, Writing - Original Draft, Writing - Review \& Editing, Visualization.
\textbf{Alberto Aleta:} Conceptualization, Writing - Original Draft, Writing - Review \& Editing.
\textbf{Yamir Moreno:} Conceptualization, Writing - Review \& Editing. Supervision, Project administration.

\section*{Funding} 
AA acknowledges support through the grant RYC2021-033226-I funded by MCIN/AEI/\\10.13039/501100011033 and the European Union ``NextGenerationEU/PRTR''. Y.M was partially supported by the Government of Aragon, Spain and ``ERDF A way of making Europe'' through grant E36-20R (FENOL), and by Ministerio de Ciencia e Innovacion, Agencia Española de
Investigacion (MCIN/AEI/10.13039/501100011033) Grant No. PID2020-115800GB-I00. 

\section*{Data Availability Statement}
The survey data is fully available and it is described in Ref~\cite{mossong} and Ref.~\cite{mistry}. All the analysis was implemented in Python 3.8.12 except when fitting the negative binomial distribution with survey data, which was implemented in R 4.2.1. The code is available upon request. 
 

\section{Figures}

\clearpage

\newpage

\section{Supplementary Methods}
\subsection{Networks}\label{supNet}
\begin{figure}[!t]
    \begin{center}
        \includegraphics[width=0.8\columnwidth]{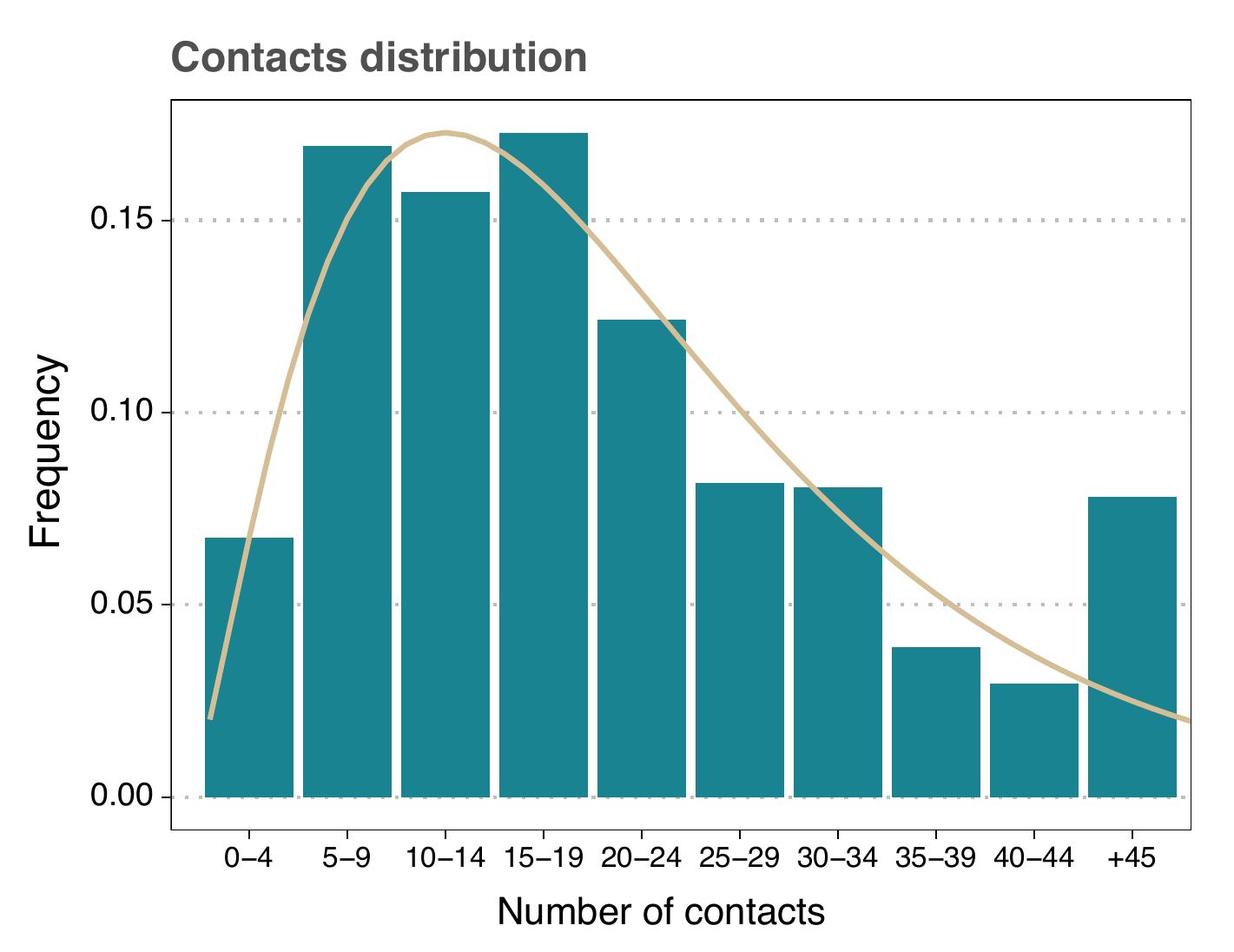}
    \end{center}
    \caption{Fit of the survey data from Ref.~\cite{mossong}. The best fit is obtained using a Negative Binomial distribution with $\mu=11.85$ and $r=2.426$.}\label{fit}
\end{figure}

Ref.\cite{mossong} identified that the degree distribution of risk contacts in an epidemic contagion scenario of an air-borne disease follows a negative binomial distribution. Thus, the degree distribution of our realistic population was assumed to follow this distribution with the shape parameters ($n$ and $p$) parameterised according to survey data. A negative binomial distribution is usually described through the probability mass function

\begin{equation}
    f(k)= \binom{k + r -1}{k}p^n (1-p)^k
\end{equation}
\noindent
where $r$ is the number of successes ($r\geq0$), $k$ is the number of failures, and $p$ is the probability of a single success. The distribution can also be parameterised in terms of the mean number of failures ($\mu$) needed to succeed \cite{cook}, by defining the probability of success ($p$) as
\begin{equation}
    p= \frac{r}{r+\mu}.
\end{equation} 
Through this definition, the distribution can be completely characterised by only fitting $r$ and $\mu$. In our case, $\mu$ is the average number of contacts of each individual, their average degree ($\langle k \rangle$). We estimated it from the age-mixing matrices for the Italian population described in Ref.\cite{mistry}. As we did not consider age heterogeneities in our analysis, the population's average degree was defined as the average number of contacts across all age groups ($\mu=11.85$). For estimating the $r$ parameter, we fit a negative binomial distribution with $\mu=11.85$ using the survey data from the POLYMOD study~\cite{mossong}. The fitting process was performed in R (version 4.2.1), using the approach proposed in Ref~\cite{lu} and it allowed us to identify that $r=2.426$ (see Figure \ref{fit}). 

To define the degree of each node we used the function \textit{nbinom.rvs()} from the Python package \textit{SciPy} (version 1.7.3) \cite{scipy} to draw 10,000 random samples from the estimated degree distribution. Nonetheless, in epidemic processes, the disease can only spread if $k>2$. To model this effect while maintaining the desired  $\langle k \rangle$ we performed the sampling process using a negative binomial distribution with 
 \begin{equation}
     \begin{alignedat}{2}
        &r &&=2.426, \\ 
        &\mu^*&&= \mu- k_{min}, \\
        &p&&=\frac{n}{n+\mu^*-k_{min}}.
     \end{alignedat}
\end{equation}
\noindent
where $k_{min}=2$. The distribution obtained has the same shape as the negative binomial extracted from the survey data but with $\mu^*= \mu-2$. In this way, we can sum $k_{min}=2$ to all the sampled values, to obtain the desired $\langle k \rangle$ while meeting the $k_{min}$ criteria. Finally, the list of degrees was transformed into a graph using the configurational model \cite{barabasi} with the Fabien Viger implementation \cite{igraph}, which creates undirected, connected, simple graphs while respecting the desired degree distribution. 

\begin{figure}[!ht]
    \begin{center}
        \includegraphics[width=0.8\columnwidth]{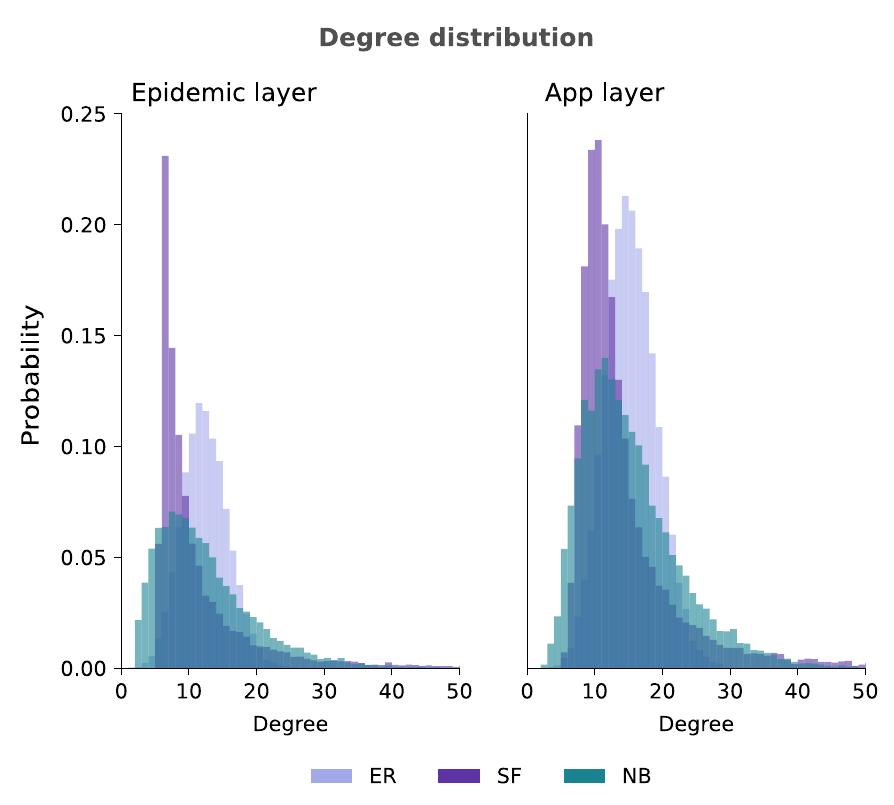}
    \end{center}
    \caption{Comparison of the degree distributions of the networks in the epidemic (left-panel) and CT app (right-panel) layers. The three distributions reflect respectively, an Erd\H{o}s-Rényi random graph (ER), a Scale-free network (SF) and a Negative Binomial degree distribution fit with survey data (NB).}~\label{distributions}
\end{figure}
To test the generality of the results we repeated the analysis using two artificial degree distributions with the same $\langle k \rangle$, an Erd\H{o}s-Rényi~\cite{erdos} random graph and a Scale-Free network~\cite{barabasi}. Erd\H{o}s-Rényi~\cite{erdos} random graphs reflect homogeneous mixing in a population. All individuals have the same probability ($p=\frac{1}{N}$) to interact with others. The Erd\H{o}s-Rényi network has a degree distribution with a binomial shape centred on the $\langle k \rangle$. The distribution of the Scale-Free network follows a power law with an exponent $\alpha_{PL}=2.5$ and the $x_{min}$ parameter was tuned to produce a distribution with the desired average degree. To generate a scale-free network with uncorrelated degree nodes, the maximal degree of the network was constrained to
\begin{equation}
     k_{max}=\sqrt{N}
\end{equation}
where $N$ is the number of individuals in the network \cite{catanzaro}. To this end, nodes with a degree $>k_{max}$ were re-sampled until the constraint was met. As for the negative binomial distribution, the connectivity pattern was also generated using the configurational model implemented in the \textit{igraph} package (version 0.9.8)~\cite{igraph}. 

The connectivity pattern in the CT app network was generated by expanding the network on top of which the epidemic spreads with 25\% of random interactions, which represent spurious interactions misinterpreted by the CT app as risk contacts. To create them, we generated a large pool of potential interactions that are not already present in the network and are not self-loops. We sampled the desired number of edges from the pool and then included them in the original network. This process was performed in the same way for the three population networks.

The degree distributions obtained in the three population structures are shown in figure \ref{distributions}. 
\clearpage
\subsection{Parameters}
Table \ref{parameters} shows the complete list of all the parameters of the epidemic model, the CT app adoption dynamics, the networks used for the analysis and the simulations. Parameters marked with a "-" are the three human behavioural parameters characterised through our analysis. 

\begin{table}[!h]
\centering \label{table_param}
\begin{tabular}{cccc}
\multicolumn{4}{c}{\textbf{PARAMETERS OF THE DYNAMICAL MODEL}} \\ \hline
\multicolumn{2}{|c||}{\textbf{Epidemic model}} & \multicolumn{2}{c|}{\textbf{CT app model}} \\ \hline
\multicolumn{1}{|c|}{$I_{0}$} & \multicolumn{1}{c||}{0.0005} & \multicolumn{1}{c|}{$App_{0}$} & \multicolumn{1}{c|}{0.01} \\ \hline
\multicolumn{1}{|c|}{$\beta$} & \multicolumn{1}{c||}{0.045} & \multicolumn{1}{c|}{$I_{thr}$} & \multicolumn{1}{c|}{-} \\ \hline
\multicolumn{1}{|c|}{$\epsilon$} & \multicolumn{1}{c||}{1/3} & \multicolumn{1}{c|}{Percentage of compliant users} & \multicolumn{1}{c|}{-} \\ \hline
\multicolumn{1}{|c|}{$\rho$} & \multicolumn{1}{c||}{1/2} & \multicolumn{1}{c|}{$\alpha$} & \multicolumn{1}{c|}{1/10} \\ \hline
\multicolumn{1}{|c|}{$\mu$} & \multicolumn{1}{c||}{1/2} & \multicolumn{1}{c|}{Max percentage of adoption} & \multicolumn{1}{c|}{-} \\ \hline
\multicolumn{1}{|c|}{Leave quarantine prob.} & \multicolumn{1}{c||}{1/10} & \multicolumn{2}{c|}{\multirow{2}{*}{}} \\ \cline{1-2}
\multicolumn{1}{|c|}{Detection rate} & \multicolumn{1}{c||}{0.5} & \multicolumn{2}{c|}{} \\ \hline
\\
\multicolumn{4}{c}{\textbf{PARAMETERS OF THE NETWORK STRUCTURE}} \\ \hline
\multicolumn{2}{|c||}{\textbf{Epidemic network}} & \multicolumn{2}{c|}{\textbf{CT app network}} \\ \hline
\multicolumn{1}{|c|}{Number of nodes} & \multicolumn{1}{c||}{10,000} & \multicolumn{1}{c|}{Number of nodes} & \multicolumn{1}{c|}{10,000} \\ \hline
\multicolumn{1}{|c|}{$\langle k_{ER} \rangle$} & \multicolumn{1}{c||}{11.93} & \multicolumn{1}{c|}{$\langle k_{ER} \rangle$} & \multicolumn{1}{c|}{14.92} \\ \hline
\multicolumn{1}{|c|}{$\langle k_{SF} \rangle$} & \multicolumn{1}{c||}{11.92} & \multicolumn{1}{c|}{$\langle k_{SF} \rangle$} & \multicolumn{1}{c|}{14.90} \\ \hline
\multicolumn{1}{|c|}{$\alpha_{PL}$} & \multicolumn{1}{c||}{2.5} & \multicolumn{1}{c|}{$\alpha_{PL}$} & \multicolumn{1}{c|}{2.5} \\ \hline
\multicolumn{1}{|c|}{$x_{min}$ SF} & \multicolumn{1}{c||}{5.3} &  \multicolumn{1}{c|}{$x_{min}$ SF} & \multicolumn{1}{c|}{10.45}\\ \hline
\multicolumn{1}{|c|}{$\langle k_{NB} \rangle$} & \multicolumn{1}{c||}{11.85} & \multicolumn{1}{c|}{$\langle k_{NB} \rangle$} & \multicolumn{1}{c|}{14.82} \\ \hline
\multicolumn{1}{|c|}{$r_{NB}$} & \multicolumn{1}{c||}{2.426} & \multicolumn{1}{c}{} & \multicolumn{1}{c|}{} \\  \hline
\multicolumn{1}{l}{} & \multicolumn{1}{l}{} & \multicolumn{1}{l}{} & \multicolumn{1}{l}{} \\ 

\multicolumn{4}{c}{\textbf{PARAMETERS OF THE SIMULATIONS}} \\ \hline
\multicolumn{2}{|c||}{time-steps (days)} & \multicolumn{2}{c|}{500} \\ \hline
\multicolumn{2}{|c||}{Repetitions} & \multicolumn{2}{c|}{1000} \\ \hline
\end{tabular}
\caption{Summary of all the parameters used in the epidemic-CT app model. The average reluctancy threshold, percentage of compliant users and maximal percentage of adoption are the quantities explored in the analysis.}\label{parameters}
\end{table}

\subsection{Alignment of repetitions}\label{align}
The 1000 repetitions of each simulation were aligned before estimating the average response and the confidence interval (CI). This ensures that the average is always estimated over an equivalent point in disease progression, regardless of the delay in the emergence of the outbreak induced by the stochastic nature of the simulations. The alignment process followed the approach described in Ref.~\cite{kiss}, where $t=0$ is defined at the time-step where all outbreaks have $1\%$ of the total population infected. This process modifies the temporal overlap of the simulations, generating some non-overlapping regions at both ends. To address it, we padded all simulations with their edge values until the full temporal range was complete.

\subsection{7-day incidence estimation}
We reported the temporal evolution of the epidemic and app adoption dynamics using the average results for the last 7 days. This process was performed by using the convolution of the incidence, prevalence and app adoption trends with a uniform window of 7 time steps (7 days). The convolution process allows us to reduce the noise on the incidence trends and to provide a more realistic representation of incidence progression, which usually is not reported daily. 
\vspace{1cm}
\section{Supplementary Results}
\subsection{Temporal evolution for different parameters}
\begin{figure}[!ht]
\begin{center}
\includegraphics[width=\columnwidth]{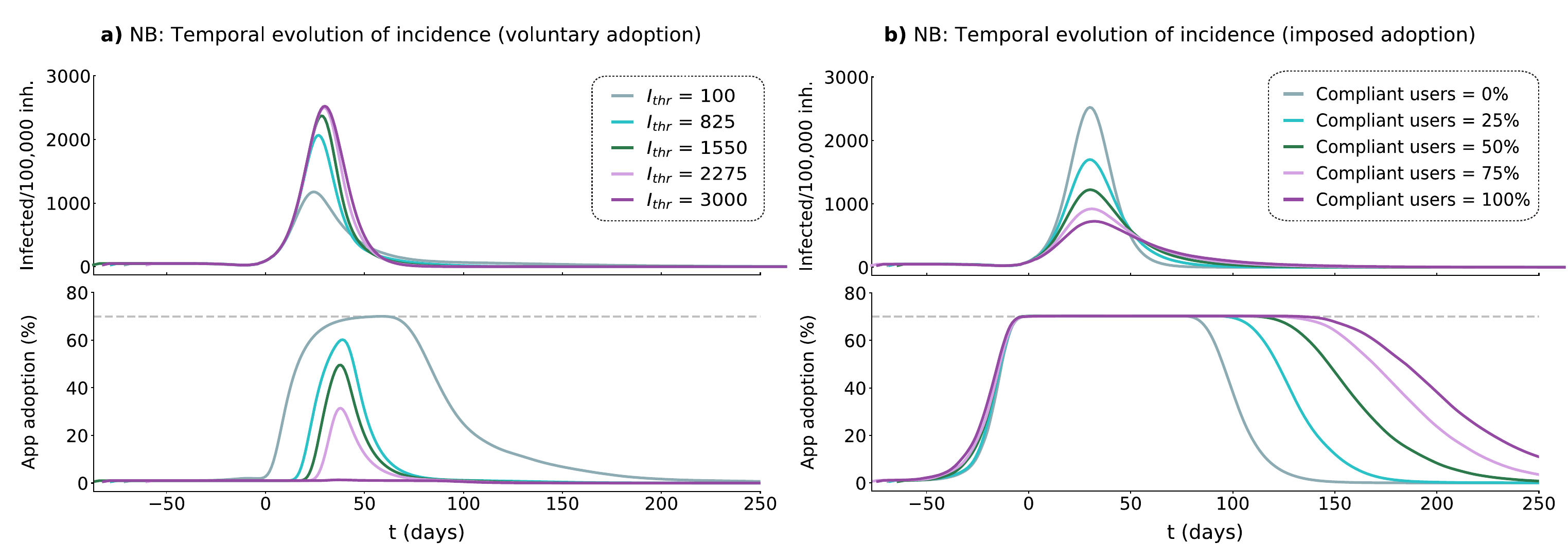}
\end{center}
\caption{Temporal evolution of the epidemic and CT app adoption dynamics for increasing values of the driving parameter of each scenario. \textbf{(A)} Changes in the incidence and the percentage of app adoption for reluctancy thresholds in the $100<I_{\text{thr}}<3000$ range. \textbf{(B)} Changes in the incidence and the percentage of app adoption for percentages of compliance ranging from 0 to 100\%.}\label{sup_traces}
\end{figure}

Figure \ref{sup_traces} shows the differences in the progression of the epidemic and CT app adoption dynamics for different reluctancy thresholds (figure \ref{sup_traces}A) and percentage of compliant users (figure \ref{sup_traces}B). We observe that CT apps introduced early in disease progression (low threshold) result in a higher peak of adoption and thus they induce a flattening of the epidemic curve. Additionally, CT apps with a lower threshold are adopted for a wider time window, also contributing to their effectiveness. Contrarily, CT apps introduced very close to the epidemic peak (or after it) almost have no effect in flattening the epidemic curve. 

Changes in compliance do not affect directly the app adoption process, they only modify the effectiveness of the reporting system. However, the coupling between disease progression and app adoption induces some indirect effects. The app adoption dynamic grows in the same way for all compliance levels, reaching almost simultaneously the maximal level of adopters (70\%). However, we do observe that modifying compliance alters the time of removal of the CT app. This indirect effect is derived from a change in the flattening of the epidemic curve produced by the CT app. In this scenario if the epidemic curve is flatter and lasts for longer,  users maintain the app downloaded for a longer period, which also affects the performance of the strategy. 

\clearpage
\subsection{Exploratory analysis of the human behavioural parameters for different network structures}
\begin{figure}[!b]
\begin{center}
\includegraphics[width=\columnwidth]{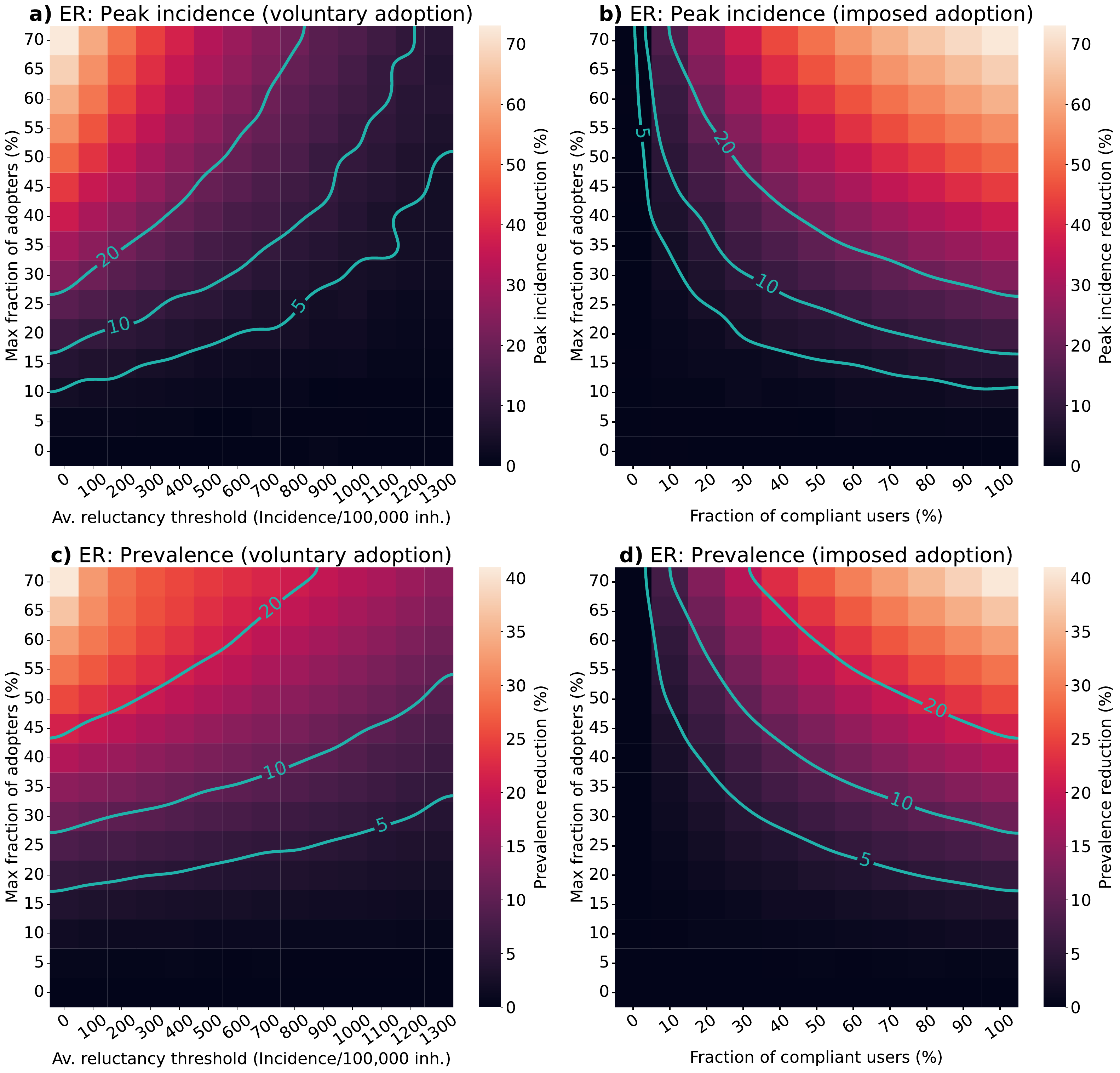}
\end{center}
\caption{Impact of different factors of human behaviour in the effectiveness of CT apps in a population with an Erd\H{o}s-Rényi random graph structure. For the voluntary adoption scenario (panels \textbf{(A)} and \textbf{(C)}), the parameters explored are the average reluctancy threshold and the maximal fraction of adopters, while in the imposed adoption scenario (panels \textbf{(B)} and \textbf{(D)}), changes in the fraction of compliant users and the maximal fraction of adopters are explored. The colour scale reflects the average reduction produced by the CT app ($\Delta$) in the peak incidence (top panels) or maximal prevalence (lower panels). The isoclines indicate the regions with $\Delta=5\%$, $\Delta=10\%$ and $\Delta=20\%$.}\label{sup_heat_ER}
\end{figure}

\begin{figure}[!t]
\begin{center}
\includegraphics[width=\columnwidth]{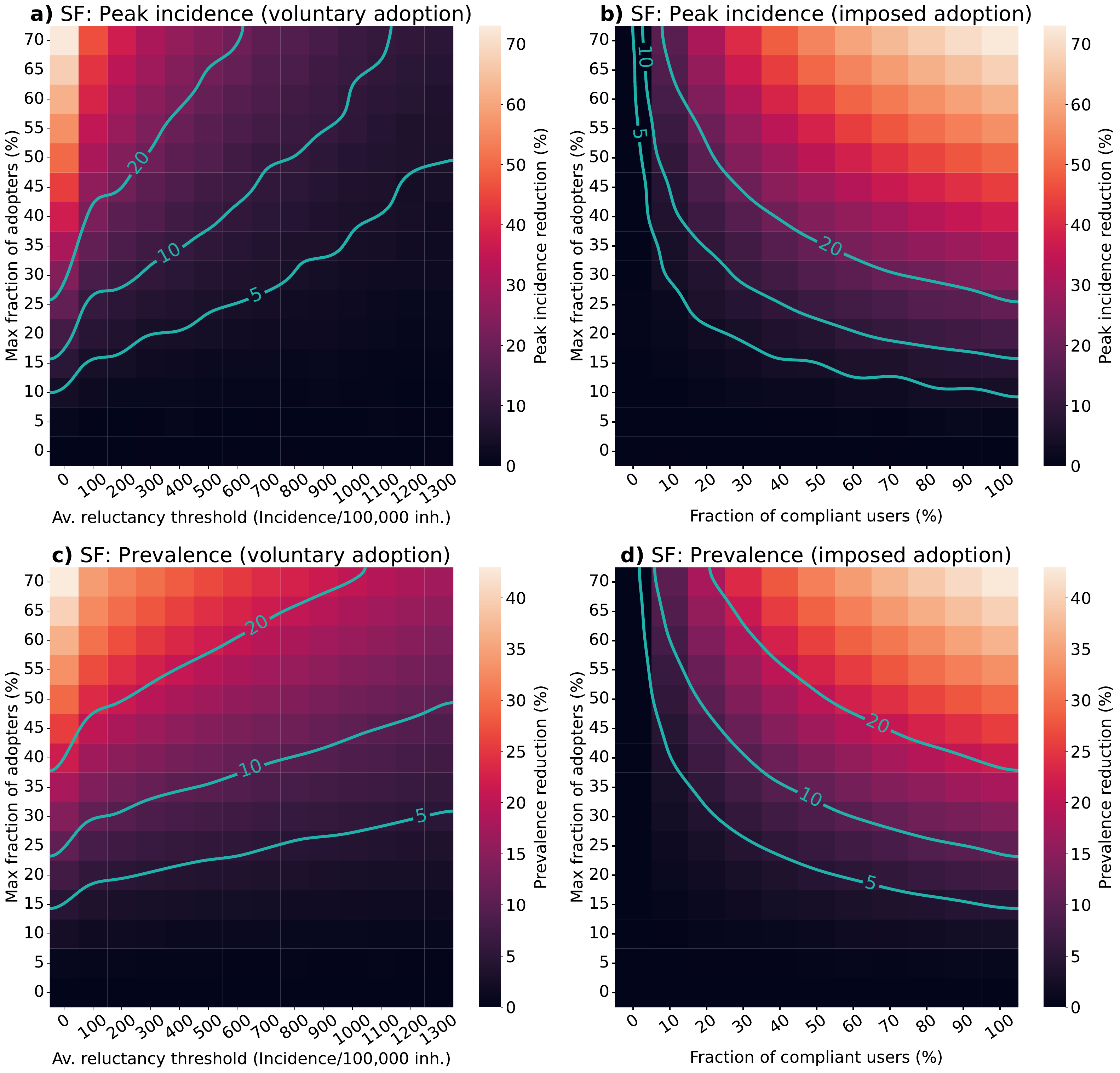}
\end{center}
\caption{Impact of different factors of human behaviour in the effectiveness of CT apps in a population with a Scale-Free degree distribution. For the voluntary adoption scenario (panels \textbf{(A)} and \textbf{(C)}), the parameters explored are the average reluctancy threshold and the maximal fraction of adopters, while in the imposed adoption scenario (panels \textbf{(B)} and \textbf{(D)}), changes in the fraction of compliant users and the maximal fraction of adopters are explored. The colour scale reflects the average reduction produced by the CT app ($\Delta$) in the peak incidence (top panels) or maximal prevalence (lower panels). The isoclines indicate the regions with $\Delta=5\%$, $\Delta=10\%$ and $\Delta=20\%$.}\label{sup_heat_SF}
\end{figure}

\begin{figure}[!ht]
\begin{center}
\includegraphics[width=\columnwidth]{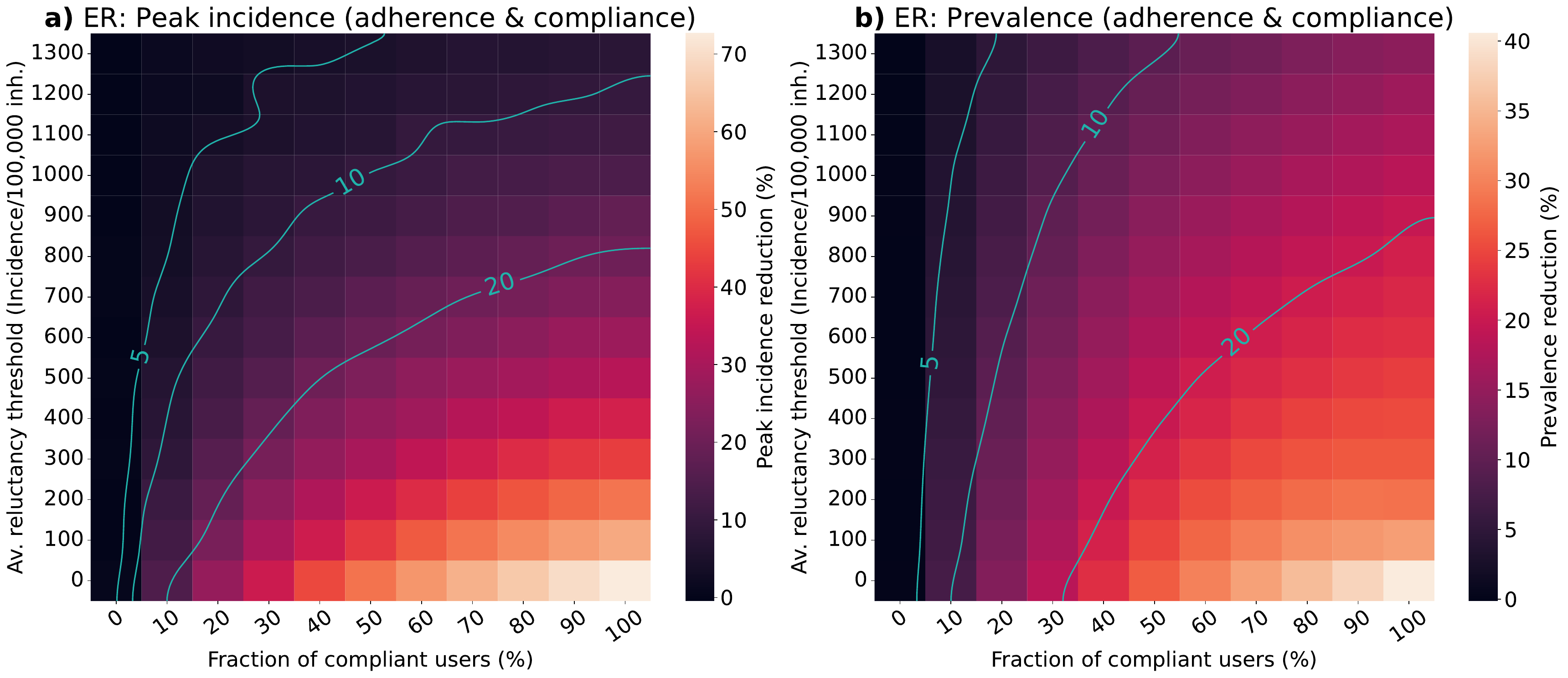}
\end{center}

\caption{Impact of the population's average reluctancy threshold and the reporting compliance in the effectiveness of CT apps for a population with an Erd\H{o}s-Rényi distribution. This scenario (``adherence \& compliance") follows the assumption that $max(App)= 70\%$. \textbf{(A)} shows the effectiveness of the CT app ($\Delta$) in terms of average peak incidence reduction, while for \textbf{(B)} the effectiveness in terms of prevalence is observed. The isoclines indicate the regions with $\Delta=5\%$, $\Delta=10\%$ and $\Delta=20\%$.}\label{ER_adcop}
\end{figure}

\begin{figure}[!ht]
\begin{center}
\includegraphics[width=\columnwidth]{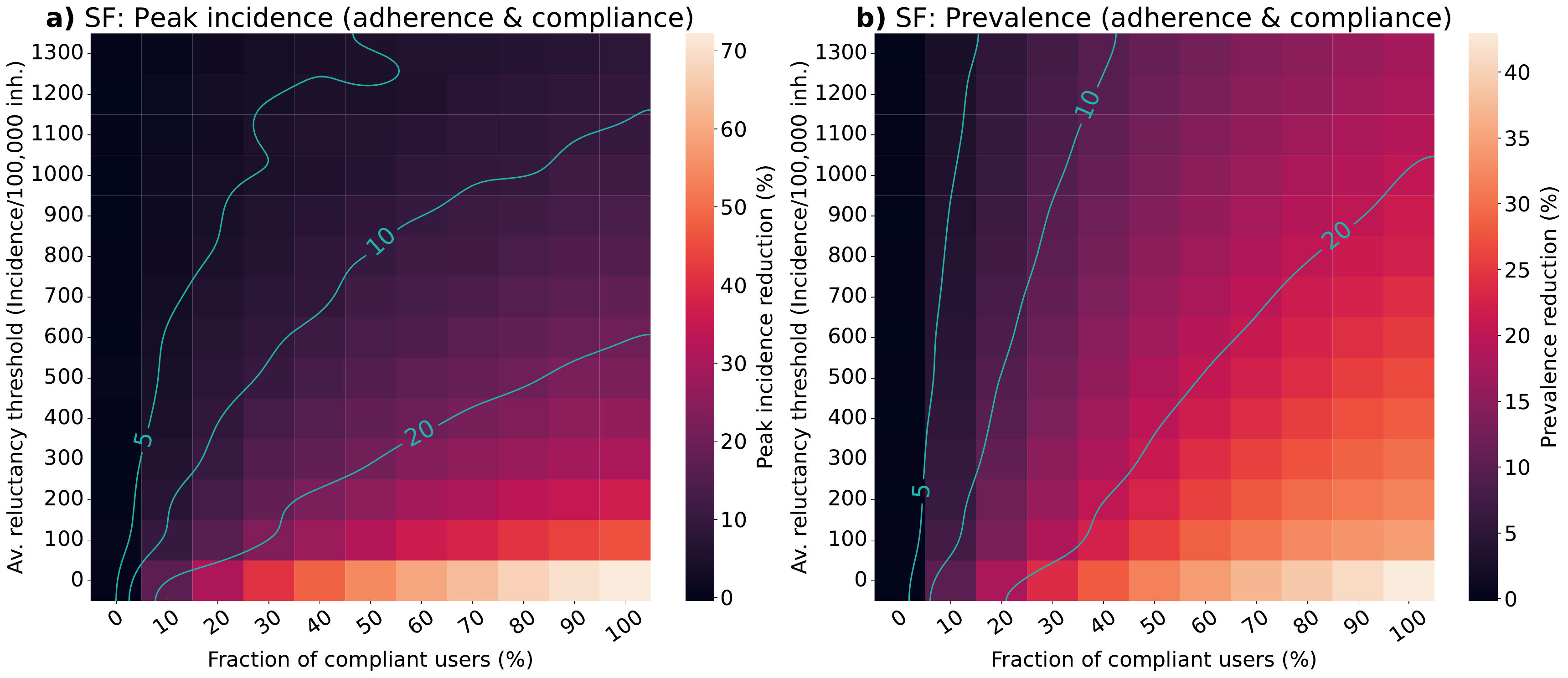}
\end{center}

\caption{Impact of the population's average reluctancy threshold and the level of compliance in the effectiveness of CT apps for a population with a Scale-Free distribution. This scenario (``adherence \& compliance") follows the assumption that $max(App)= 70\%$. \textbf{(A)} shows the effectiveness of the CT app ($\Delta$) in terms of average peak incidence reduction, while for \textbf{(B)} the effectiveness in terms of prevalence is observed. The isoclines indicate the regions with $\Delta=5\%$, $\Delta=10\%$ and $\Delta=20\%$.}\label{SF_adcop}
\end{figure}

Figure \ref{sup_heat_ER} and figure \ref{sup_heat_SF} show the exploratory analysis for the voluntary and imposed adoption scenarios in a population with an Erd\H{o}s-Rényi and a Scale-Free degree distributions, respectively. The equivalent results for the ``adherence \& compliance" scenario can be found in figure~\ref{ER_adcop} and figure~\ref{SF_adcop}. The results obtained are consistent with the ones for the realistic population (negative binomial distribution). High adoption and moderate levels of compliance are crucial for effective strategies. Additionally, we observe that the random population has a very similar profile to the negative binomial network, with a slightly wider effective parameter space than in the realistic case. Contrarily, the Scale-free population shows a very narrow parameter space with effective apps. This may result from the increased transmissibility in the heterogeneous population, which diminishes the effectiveness of the strategy. 
\vfill

\newpage
\null
\vfill
\vfill

\clearpage



\end{document}